\newcommand\araa{{ARA\&A\,}}%
\newcommand\apj{{ApJ\,}}%
\newcommand\apjl{{ApJ\,}}%
\newcommand\aap{{A\&A\,}}%
\newcommand\mnras{{MNRAS\,}}%
\newcommand\pasp{{PASP\,}}%
\newcommand\nat{{Nature\,}}%

\def\Dnu{$\Delta\nu$}
\def\numax{$\nu_{\rm max}$}
\def\kepler{\mbox{\textit{Kepler}}}

\documentclass[graybox, natbib, footinfo]{svmult}

\usepackage{mathptmx}       
\usepackage{helvet}         
\usepackage{courier}        
\usepackage{type1cm}        
\usepackage{makeidx}         
\usepackage{graphicx}        
\usepackage{multicol}        
\usepackage[bottom]{footmisc}

\usepackage{lscape}
\usepackage{amssymb}

\makeindex             

\begin{document}

\title*{Solar-like oscillating stars as standard clocks and rulers for Galactic studies}
\author{Andrea Miglio,  L\'eo Girardi, Tha\'{\i}se S. Rodrigues, Dennis Stello \& William J. Chaplin}
\authorrunning{Miglio et al.} 
\institute{A. Miglio, W. J. Chaplin \at School of Physics and Astronomy, University of Birmingham, United Kingdom\\
Stellar Astrophysics Centre (SAC), Department of Physics and Astronomy, Aarhus University
\and L. Girardi, T. S. Rodrigues \at INAF -- Osservatorio Astronomico di Padova, Vicolo dell'Osservatorio 5, Padova, Italy
\and
D. Stello \at Sydney Institute for Astronomy (SIfA), School of Physics, University of Sydney, Australia\\
Stellar Astrophysics Centre (SAC), Department of Physics and Astronomy, Aarhus University
}
%
%
\maketitle

\abstract{The CoRoT and {\it Kepler} space missions have detected oscillations in hundreds of Sun-like stars and thousands of field red-giant stars. This has opened the door to a new era of stellar population studies in the Milky Way. We report on the current status and future prospects of harvesting space-based photometric data for ensemble asteroseismology, and highlight some of the challenges that need to be faced to use these stars as accurate clocks and rulers for Galactic studies.}

\section{Introduction}
Prior to the advent of space-based photometric missions such as \kepler\ and CoRoT, age estimation of stars had to rely on constraints from their surface properties only. A severe limitation of such an approach applies in particular to objects in the red-giant phase: that stars of significantly different age and distance end up sharing very similar observed surface properties making it extremely hard to, for example, map and date giants belonging to the Galactic-disc population. That situation has now changed thanks to asteroseismology, with the detection of solar-like pulsations in hundreds of dwarfs and thousands of red giants observed by CoRoT and \kepler. The pulsation frequencies may be used to place tight constraints on the fundamental stellar properties, including radius, mass, evolutionary state, internal rotation and age (see e.g. \citealt{Chaplin2013} and references therein).

There are several important reasons why asteroseismic data, and in particular data on red
giants, offer huge potential for populations studies. First, G and K
giants are numerous. They are therefore substantial contributors by
number to magnitude-limited surveys of stars, such as those conducted by CoRoT and
\kepler. Moreover, the large intrinsic oscillation amplitudes and long
oscillation periods mean that oscillations may be detected in faint
targets observed in the ``standard'' long-cadence modes of operation of both CoRoT and \kepler. Second, with asteroseismic data in hand red
giants may be used as accurate distance indicators probing regions out
to about $10\,\rm kpc$. As in the case of eclipsing binaries, the
distance to each red giant may be estimated from the absolute
luminosity, which is obtained from the asteroseismically determined
radius and $T_{\rm eff}$. This differs from the approach adopted to
exploit pulsational information from classical pulsators, notably
Cepheid variables, where the observed pulsation frequency leads to an
estimate of the mean density only and hence additional calibrations and
assumptions are needed to yield an estimated distance.

Third, seismic data on red-giant-branch (RGB) stars in principle provide robust ages that
probe a wide age range. Once a star has evolved to the RGB, its age is
determined to a first approximation by the time spent in the
core-hydrogen burning phase, which is predominantly a function of mass
and metallicity. Hence, the estimated masses of red giants provide
important constraints on age. The CoRoT and \kepler\, giants cover a mass
range from $\simeq 0.9$ to $\simeq 3\,\rm M_\odot$, which in turn maps
to an age range spanning $\sim0.3$ to $\sim 12\,\rm Gyr$, i.e., the
entire Galactic history. 

Sun-like stars with detectable oscillations span a much more limited distance than the giants. When individual oscillation frequencies are available, however, the age determination of such stars can rely  on the use of diagnostics specific to the conditions of the star's core, which can be used as proxies for the evolutionary state on the main sequence, and hence the absolute age  \citep[e.g.][]{JCD88, Roxburgh03}. 

We are still in the infancy of using solar-like oscillators as probes of the Milky Way's properties, and the field is evolving so rapidly that any attempts at reviewing the status would be quickly outdated. In what follows we will therefore briefly summarise the asteroseismic data currently available, and which data will be available in the next decade, and flag some of the challenges that need to be faced in {\it ensemble asteroseismology}, i.e. the study of stellar populations including constraints from global, resonant oscillation modes.

\begin{figure*}
\centering
   \includegraphics[width=.5\hsize, angle=-90]{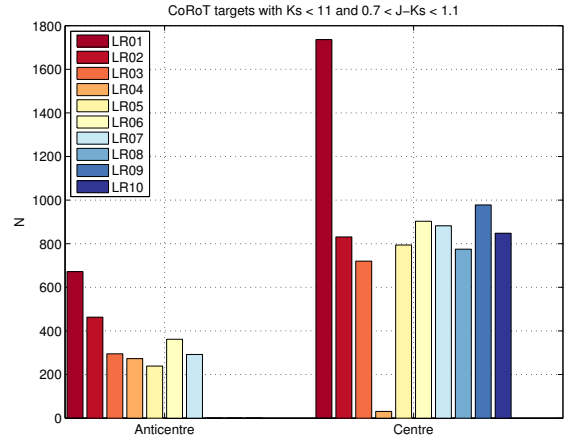}
      \caption{Number of stars observed in CoRoT's exofields in the colour-magnitude range expected to be populated by red giants with detectable oscillations ($0.7<J-K_s<1.1, K_s < 11$). The analysis of the first two observational runs led to the detection of oscillations in about 1600 (LRc01) and 400 (LRa01) stars, as reported in \cite{Hekker09} and  \citet{Mosser2010}. The varying number of stars per field reflects different target selection functions used, the failure of 2 CCD modules, as well as the different stellar density in each field.}
         \label{fig:corotfields}
\end{figure*}

\section{Harvesting data for ensemble seismology}
The full exploitation of the available CoRoT and \kepler\  asteroseismic data is far from complete, both in terms of the extraction of seismic parameters and the characterisation of the observed stellar populations.

\begin{figure*}
\centering
   \includegraphics[width=.6\hsize, angle=0]{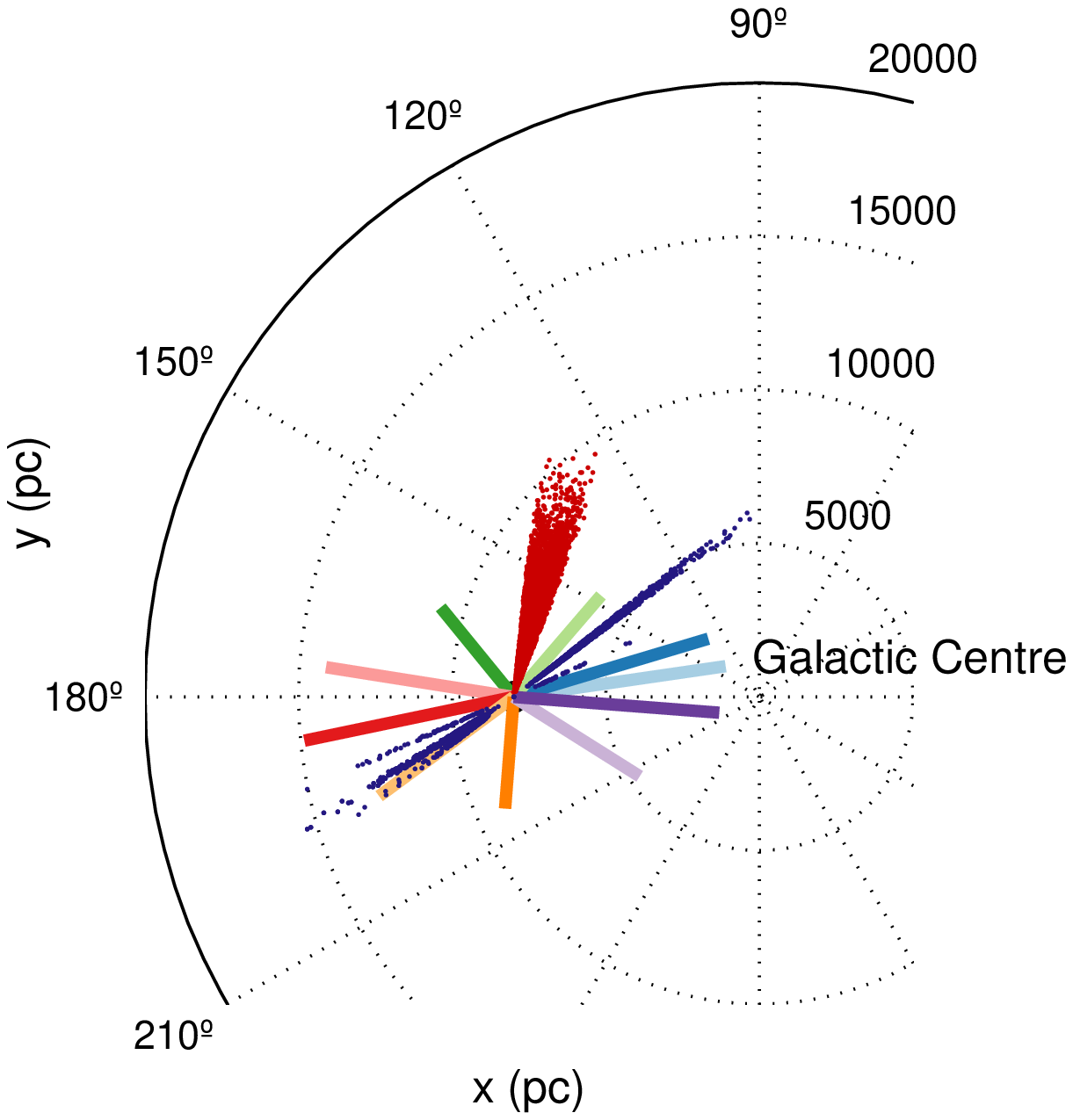}
   \includegraphics[width=.8\hsize, angle=0]{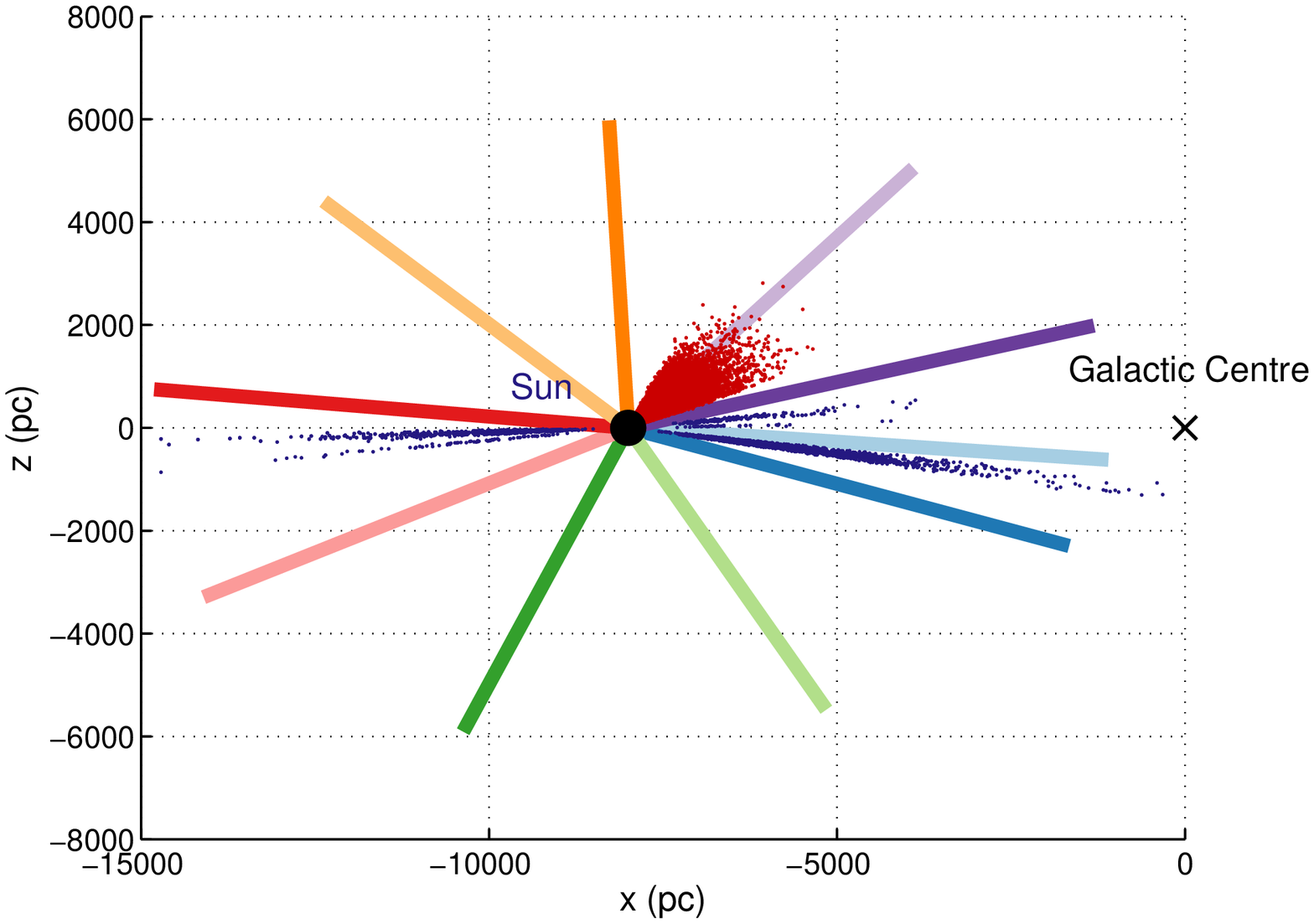}
   \includegraphics[width=1\hsize, angle=0]{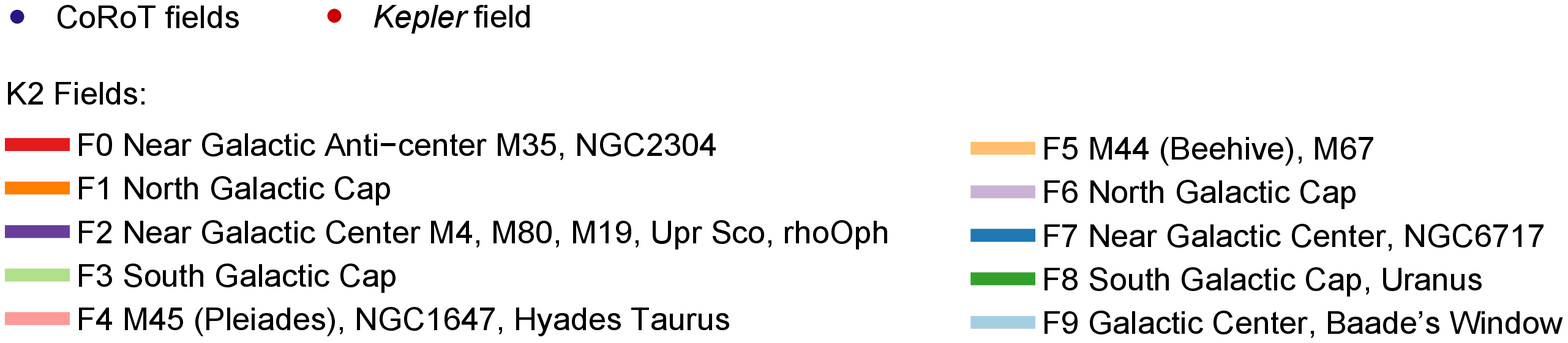}

      \caption{Location in the Galaxy of stars with seismic constraints observed by CoRoT, \kepler\, and in the fields which will be monitored by K2.}
         \label{fig:fields}
\end{figure*}

Only two CoRoT long-observational runs have been analysed so far \citep{Hekker09, Mosser2011, Miglio2013}. Data collected in 15 additional runs, crucially exploring stellar populations at different galactocentric radii, are yet to be exploited. Figure \ref{fig:corotfields} shows the number of stars targeted in CoRoT's observational campaigns in the colour-magnitude range expected to be populated by red giants with detectable oscillations. The analysis of these data is currently ongoing.

The full mining of the ``nominal mission'' \kepler\ data is also in its early stages. The scientific community is now just beginning to tackle the detailed fitting of  individual oscillation modes. Multi-year data are now available on about one hundred solar-type stars and $\sim 20,000$ red giants \citep{Hekker2011b, Stello2013}.

The outlook for the collection of new data in the near future also looks remarkably bright. In addition to CoRoT and \kepler, three more missions will supply asteroseismic data for large samples of stars: the re-purposed \kepler\ mission, K2 \citep{Howell2014} and, in a few year's time, the space missions TESS\footnote{\tt http://tess.gsfc.nasa.gov/} \citep{Ricker2014} and PLATO 2.0\footnote{\tt http://sci.esa.int/plato} \citep{Rauer2013}.

Although \kepler\ ceased normal operations in June 2013, we can now look forward to a new mission using the \kepler\ spacecraft and instrument. K2 has started performing a survey of stars in the ecliptic plane, with each pointing lasting around 80\,days. The potential of  K2 to study stellar populations is even greater than for CoRoT and \kepler, since regions of the sky near the ecliptic contain bright clusters and will also make it possible to map the vertical and radial structure of the Galaxy (see Fig. \ref{fig:fields}). The importance of this line of research was acknowledged by it being listed as a potential mission highlight by \citet{Howell2014}.  K2 was recently  approved by the NASA senior review, and observations of $\sim8,600$ giants in the first full science campaign has already been made, and $5,000$ more will be observed in Campaign 2. Future K2 observations may also provide constraints on giants belonging to the Galactic bulge (Campaign 9) and, crucially, in the globular cluster M4 which would serve as a much needed test for seismic scaling relations in the metal poor regime \citep{Epstein2014} .


\begin{figure*}
\centering
   \includegraphics[width=.8\hsize, angle=0]{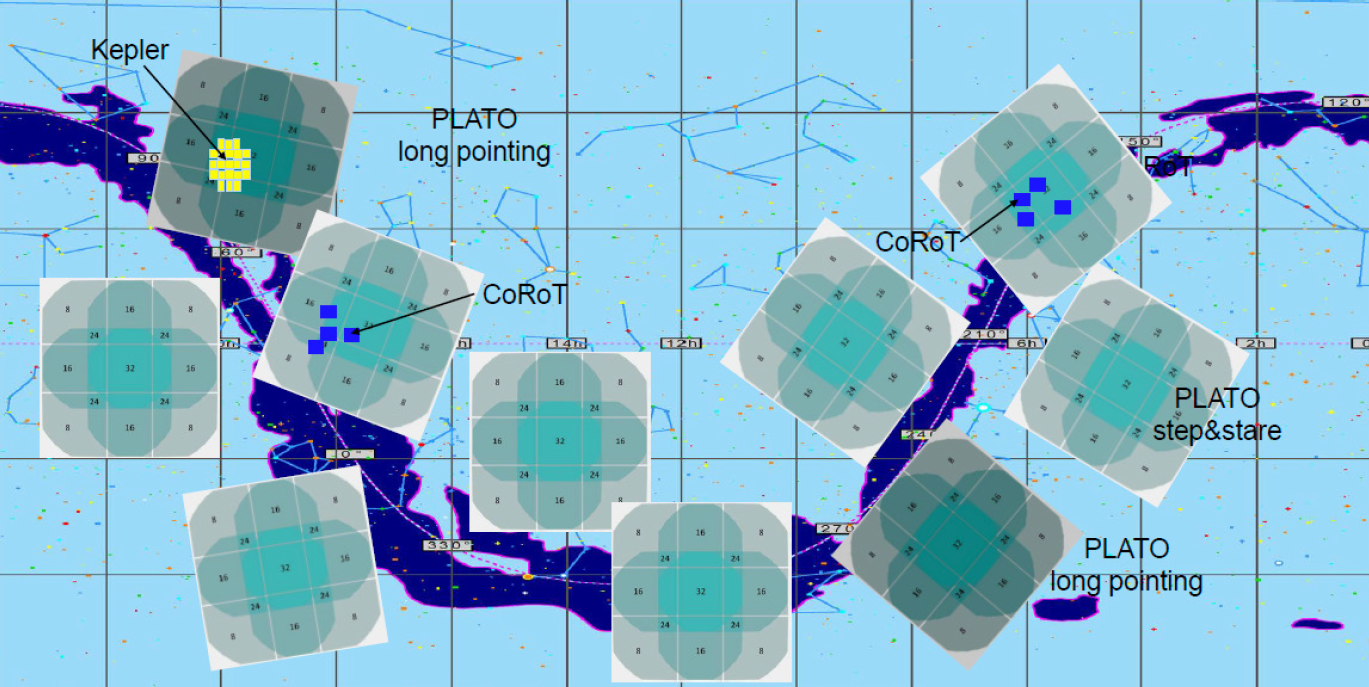}
      \caption{Schematic comparison of PLATO 2.0, CoRoT and \kepler's fields of view and observational strategy. With a combination of short (step-and-stare) and long duration pointings PLATO 2.0 will cover a large fraction of the sky. Note that the final locations of long and step-and-stare fields will be defined after mission selection and are drawn here for illustration only. Figure taken from {\tt http://www.oact.inaf.it/plato/PPLC/Science.html}.}
         \label{fig:plato}
\end{figure*}

The all-sky mission TESS \citep{Ricker2014} will also significantly add to the harvest of bright Sun-like stars in the solar neighborhood, and is expected to be launched in 2017.
Thanks to the early achievements with CoRoT and \kepler, Galactic archeology has now been included as one of the major scientific goals of ESA's medium-class mission PLATO 2.0 (see \citealt{Rauer2013}), which will supply seismic constraints for stars over a significantly larger fraction of the sky (and volume) compared to CoRoT, \kepler, and K2.  
An illustration of PLATO 2.0's observational strategy is presented in Figure \ref{fig:plato}. By detecting solar-like oscillations in $\sim$85,000 nearby dwarfs and an even larger number of giants, PLATO 2.0 will provide a revolutionary complement to Gaia's view of the Milky Way. PLATO 2.0 is planned for launch by 2024.


%

\section{From precise to accurate clocks and rulers}
Seismic data analysis and interpretation techniques have undergone a rapid and considerable development in the last few years. However, they still suffer from limitations, e.g.:
\begin{itemize}
\item determination of individual oscillations mode parameters has been carried out for a limited set of Sun-like stars, and for only a handful of red giants;
\item stellar mass and radius estimates in most cases are based on approximated scaling relations of average seismic properties (\Dnu\ and \numax), under-exploiting the information content of oscillation modes; and:
\item systematic uncertainties on the inferred stellar properties due to limitations of current stellar models have not yet been quantified. This is crucial for age estimates, which are inherently model dependent.
\end{itemize}

Here, we mention some of the key sources of uncertainty on current estimates of stellar age and how, in some cases, we hope to make progress. We   focus on evolved stars, and refer to \citet{Lebreton2014}, and references therein, for the case of main-sequence stars.

The age of low-mass red-giant stars is largely determined by the time spent on the main sequence, hence by the mass of the red giant's progenitor. Knowledge of the star's metallicity is also key to determining the age, but we will assume in the following that metallicities will be available thanks to spectroscopic constraints (see the contributions by Morel, Johnson, Epstein, Plez \& Grevesse). With data on solar-like oscillations we can indeed estimate the mass of giants. What is challenging, however, is that if we aim at determining ages to 30\% or better, then we need to be able to the determine masses with an accuracy better than 10\%. 
Testing the accuracy of the asteroseismic mass scale to 10\% or better is very much ``work in progress''.

%

An example of possible systematic biases concerning the mass determination 
are departures from a simple scaling of \Dnu\  with the square root of the stellar mean density  \citep[see e.g.][]{White2011,Miglio2012, Miglio2013, Belkacem2013}. 
Suggested corrections to the \Dnu\ scaling are likely to depend (to a level of few percent) on the the stellar structure itself. Moreover, the average \Dnu\ is known to be affected (to a level of $\sim 1\%$ in the Sun) by our inaccurate modelling of near-surface layers.
\begin{figure*}
\centering
   \includegraphics[width=.7\hsize, angle=0]{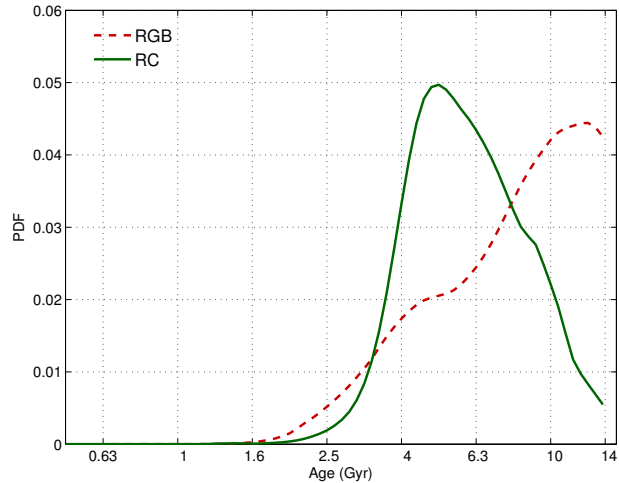}
      \caption{Example of probability distribution function of age for a star with $M=0.9$ M$_\odot$ and $R=10$ R$_\odot$ which is assumed to be either in the red-giant-branch phase (dashed line) or in the core-He burning phase (solid line). Stellar evolutionary tracks used to determine the age assume that a significant mass loss occurs following Reimers' \citep{Reimers1975a} prescription and a mass-loss efficiency parameter $\eta=0.4$.}
         \label{fig:mloss}
\end{figure*}

A  way forward would be to determine the star's mean density by using the full set of observed acoustic modes, not just their average frequency spacing. 
This approach was carried out in at least two RGB stars \citep{Huber2013, Lillo-Box2014}, and led to determination of the stellar mean density which is $\sim 5-6\%$ higher than derived from assuming scaling relations, and  with a much improved precision of  $\sim 1.4\%$.

While a relatively simple mass-age relation is expected for RGB stars,  the situation for red-clump RC or early asymptotic-giant-branch (AGB) stars is of course different.  If stars undergo a significant mass loss near the tip of the RGB, then the mass-age relation is not unique (for a given composition and input physics), since the mass observed at the RC or early-AGB stage may differ from the initial one. An extreme case is depicted in Fig. \ref{fig:mloss}, where we compare the posterior probability density function of age for a synthetic star of 0.9 M$_\odot$ which we assume to be either on the RGB or in the RC. If significant mass loss occurs, the 0.9 M$_\odot$  RGB star would have to be significantly older than if it were on in the RC phase.  In this context, the characterisation of populations of giants will benefit
greatly from estimations of the period spacings of the observed gravity
modes, which  allows a clear distinction to be made
between RGB and RC stars \citep{Bedding2011}, and early-AGB stars \citep{Montalban2013a}.
Knowledge of the efficiency of mass loss is however still needed to determine accurate ages of red/clump stars.

\begin{figure}
\hspace{-1.5cm}
   \includegraphics[width=.8\hsize, angle=-90]{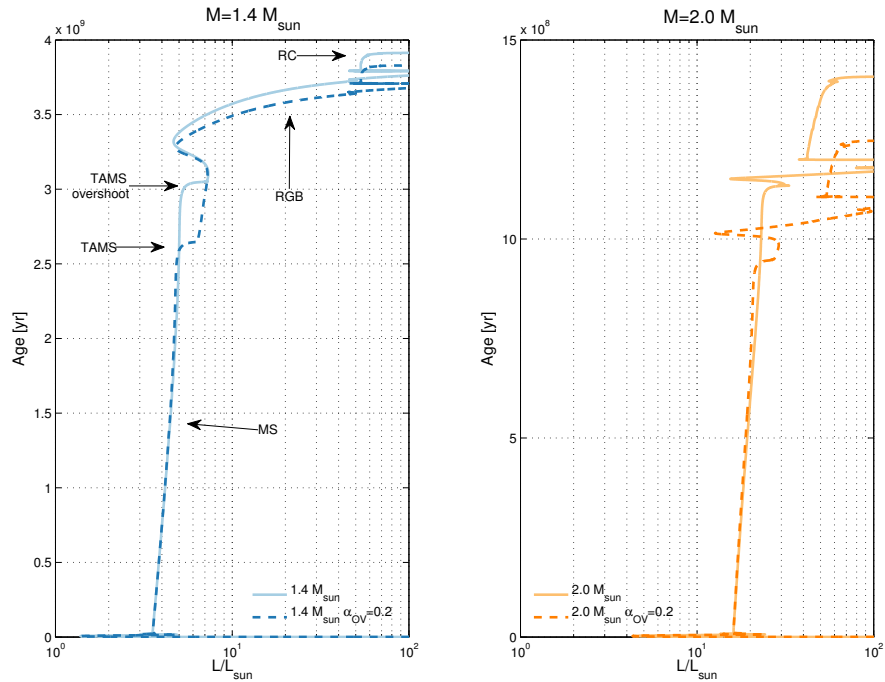}
      \caption{{\it Left panel:} Age as a function of luminosity for a 1.4 M$_\odot$ stellar model computed with (solid line) and without (dashed line) extra mixing from the convective core during the main-sequence phase. Relevant evolutionary phases are marked on the plot: MS: main sequence, TAMS: terminal-age main sequence, RGB: red-giant branch, RC: red clump. {\it Right panel:} Same as left panel but for models of a 2 M$_\odot$  star.}
         \label{fig:overshoot}
\end{figure}

Other uncertainties on the input physics may affect main-sequence lifetimes, hence the age of red giants (see Noels \& Bragaglia, this volume).
As an example, consider the impact on RGB ages of uncertainties in
predictions of the size of the central, fully-mixed region in
main-sequence stars.  We take the example of a model of mass $1.4\,\rm
M_\odot$. The difference between the main-sequence lifetime of a model
with and without overshooting\footnote{We assume an extension of the
  overshooting region equal to $0.2\,H_p$, where $H_p$ is the pressure
  scale height at the boundary of the convective core, as defined by
  the Schwarzschild criterion.} from the core is of the order of
$20\,\%$. However, once the model reaches the giant phase, this
difference is reduced to about $5\,\%$ (see Fig. \ref{fig:overshoot}). Low-mass models with a larger
centrally mixed region experience a significantly shorter subgiant
phase, the reason being that they end the main sequence with an
isothermal helium core which is closer to the Sch\"onberg-Chandrasekar
limit \citep[see][]{Maeder1975}, hence partially offsetting the impact
of a longer main-sequence lifetime.  On the other hand, the effect of
core overshooting on the age of RGB stars is more pronounced when the
mass of the He core at the end of the main sequence is close to (or
even larger than) the Sch\"onberg-Chandrasekar limit (e.g., see the  case
of a 2.0 $\rm
M_\odot$ illustrated in Fig. \ref{fig:overshoot}). In that case, however, seismology may come to the rescue as at the beginning of the core-He burning phase the period spacing of gravity modes is a proxy for the mass of the helium core, and can potentially help to test models of extra mixing for stars in the so-called secondary clump \citep{Montalban2013}.

Additional seismic diagnostics are still to be fully utilised and their dependence on stellar properties understood. Seismic signatures of sharp-structure variations can potentially lead to estimates of the envelope He abundance \citep[see ][]{Broomhall2014}. Promising indicators of global stellar properties include the small separation between radial and quadrupole modes  \citep{Montalban2012} and the properties of mixed modes and coupling term which may lead to additional indirect constraints on the stellar mass \citep[see e.g.][]{Benomar2013}.

\begin{figure*}
\centering
   \includegraphics[width=.6\hsize, angle=0]{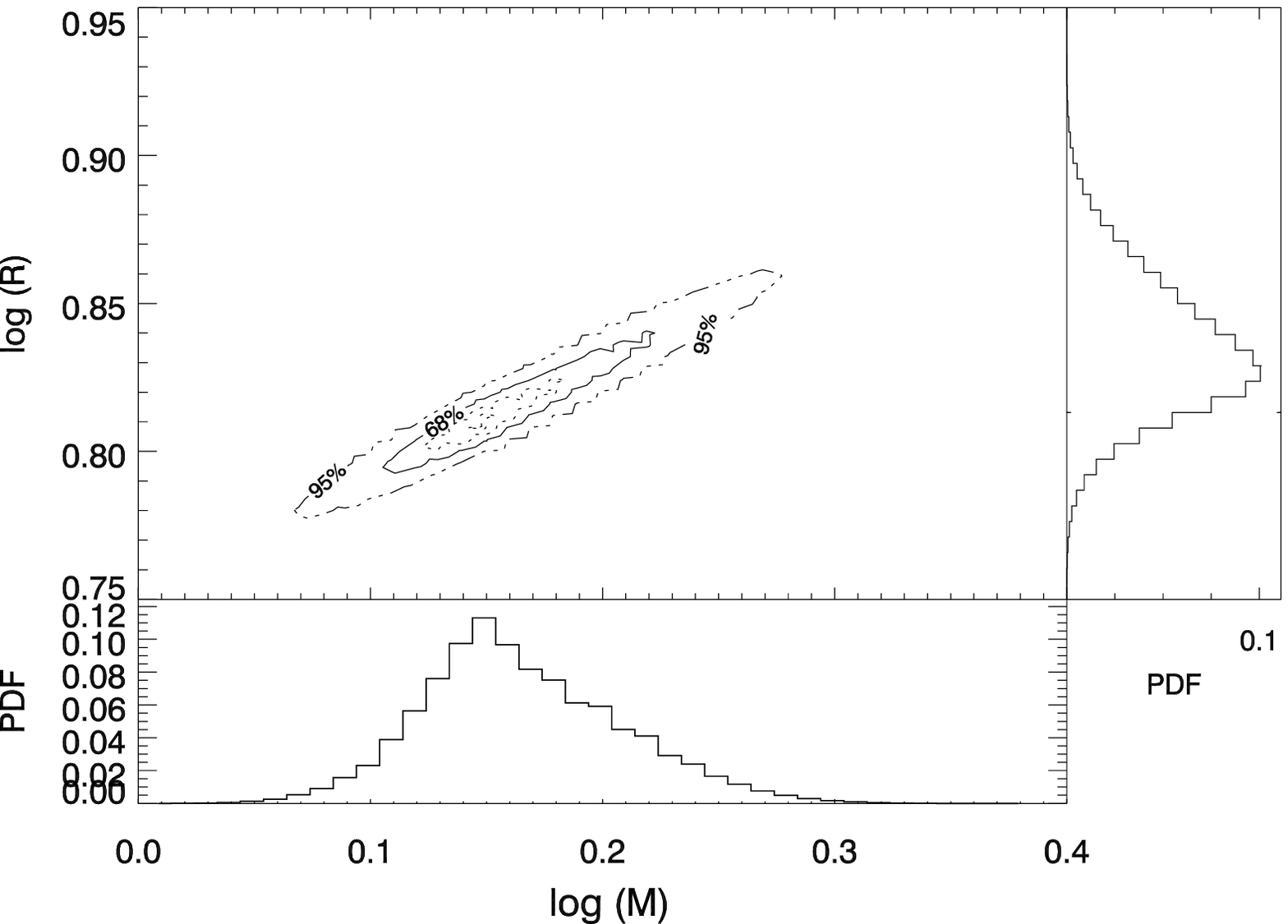}
   \includegraphics[width=.6\hsize, angle=0]{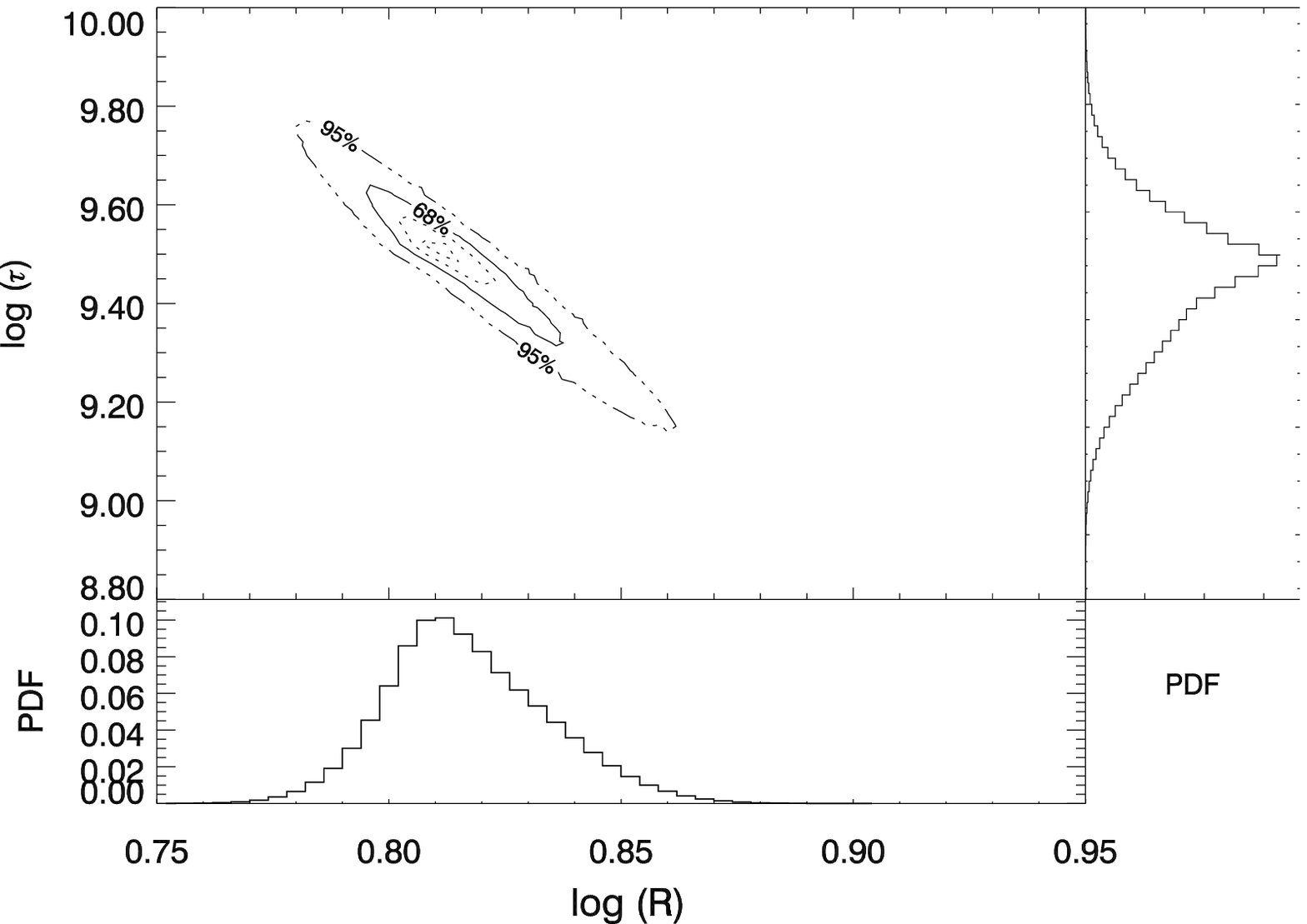}
      \caption{Contour plots of joint and marginal probability density function of radius and mass ({\it upper panel}) as output of PARAM \citep{Rodrigues2014}. The solid and dashed-dotted contours represent 68\% and 95\% credible regions. The strong correlation between mass and radius is reflected in an anticorrelation between radius (or distance) and age ({\it lower panel}).}
         \label{fig:corr}
\end{figure*}

An often overlooked point concerns the correlations between stellar properties derived by including seismic measurements, correlations which are yet to be quantified. As an example we show in Fig. \ref{fig:corr} the strong correlation between radius (hence distance) and mass (hence age) as determined by considering  average seismic quantities, effective temperature, and $\rm [Fe/H]$ as observables.

\section{Summary}
Thanks to K2, TESS, and PLATO 2.0 the future looks extremely bright in terms of collecting space-based photometric data for ensemble seismology, however, we are still faced with a number of challenges centred around the assessment of the accuracy by which the properties of stellar populations can be determined.

Robust predictions from stellar models are key to determining accurate stellar properties such as mass, radius, surface gravity and, crucially, age. A critical appraisal of how numerical and systematic uncertainties in model predictions impact the inferred stellar properties (in particular ages) is needed. 
In favourable cases (such as binary systems, clusters) stellar models will be tested against the seismic measurements and reduce (some of) the systematic uncertainties in the age determination related to, for example, near-core extra mixing during the main sequence, and mass loss on the red-giant branch.
Given the additional constraints (stringent priors on age, chemical composition) stars in clusters (see the contributions by Salaris and Brogaard et al.) and binary systems (eclipsing and seismic, see e.g. \citealt{Huber2014, Miglio2014}) represent the prime targets for testing models.

A significant step forward in ensemble seismology, as shown in this workshop,  will be through an associative development of  asteroseismic data analysis techniques, stellar modelling, spectroscopic anlyses and their combination to constrain and test chemodynamical models of the Milky Way.

%

\begin{thebibliography}{30}
\expandafter\ifx\csname natexlab\endcsname\relax\def\natexlab#1{#1}\fi

\bibitem[{{Bedding} {et~al.}(2011){Bedding}, {Mosser}, {Huber},
  {Montalb{\'a}n}, {Beck}, {Christensen-Dalsgaard}, {Elsworth},
  {Garc{\'{\i}}a}, {Miglio}, {Stello}, {White}, {De Ridder}, {Hekker}, {Aerts},
  {Barban}, {Belkacem}, {Broomhall}, {Brown}, {Buzasi}, {Carrier}, {Chaplin},
  {di Mauro}, {Dupret}, {Frandsen}, {Gilliland}, {Goupil}, {Jenkins},
  {Kallinger}, {Kawaler}, {Kjeldsen}, {Mathur}, {Noels}, {Aguirre}, \&
  {Ventura}}]{Bedding2011}
{Bedding}, T.~R., {Mosser}, B., {Huber}, D., {et~al.} 2011, \nat, 471, 608

\bibitem[{{Belkacem} {et~al.}(2013){Belkacem}, {Samadi}, {Mosser}, {Goupil}, \&
  {Ludwig}}]{Belkacem2013}
{Belkacem}, K., {Samadi}, R., {Mosser}, B., {Goupil}, M.-J., \& {Ludwig}, H.-G.
  2013, in Astronomical Society of the Pacific Conference Series, Vol. 479,
  Astronomical Society of the Pacific Conference Series, ed. H.~{Shibahashi} \&
  A.~E. {Lynas-Gray}, 61

\bibitem[{{Benomar} {et~al.}(2013){Benomar}, {Bedding}, {Mosser}, {Stello},
  {Belkacem}, {Garcia}, {White}, {Kuehn}, {Deheuvels}, \&
  {Christensen-Dalsgaard}}]{Benomar2013}
{Benomar}, O., {Bedding}, T.~R., {Mosser}, B., {et~al.} 2013, \apj, 767, 158

\bibitem[{{Broomhall} {et~al.}(2014){Broomhall}, {Miglio}, {Montalb{\'a}n},
  {Eggenberger}, {Chaplin}, {Elsworth}, {Scuflaire}, {Ventura}, \&
  {Verner}}]{Broomhall2014}
{Broomhall}, A.-M., {Miglio}, A., {Montalb{\'a}n}, J., {et~al.} 2014, \mnras,
  440, 1828

\bibitem[{{Chaplin} \& {Miglio}(2013)}]{Chaplin2013}
{Chaplin}, W.~J. \& {Miglio}, A. 2013, \araa, 51, 353

\bibitem[{{Christensen-Dalsgaard}(1988)}]{JCD88}
{Christensen-Dalsgaard}, J. 1988, in IAU Symposium, Vol. 123, Advances in
  Helio- and Asteroseismology, ed. J.~{Christensen-Dalsgaard} \& S.~{Frandsen},
  295

\bibitem[{{Epstein} {et~al.}(2014){Epstein}, {Elsworth}, {Johnson}, {Shetrone},
  {Mosser}, {Hekker}, {Tayar}, {Harding}, {Pinsonneault}, {Silva Aguirre},
  {Basu}, {Beers}, {Bizyaev}, {Bedding}, {Chaplin}, {Frinchaboy},
  {Garc{\'{\i}}a}, {Garc{\'{\i}}a P{\'e}rez}, {Hearty}, {Huber}, {Ivans},
  {Majewski}, {Mathur}, {Nidever}, {Serenelli}, {Schiavon}, {Schneider},
  {Sch{\"o}nrich}, {Sobeck}, {Stassun}, {Stello}, \& {Zasowski}}]{Epstein2014}
{Epstein}, C.~R., {Elsworth}, Y.~P., {Johnson}, J.~A., {et~al.} 2014, \apjl,
  785, L28

\bibitem[{{Hekker} {et~al.}(2011){Hekker}, {Gilliland}, {Elsworth}, {Chaplin},
  {De Ridder}, {Stello}, {Kallinger}, {Ibrahim}, {Klaus}, \&
  {Li}}]{Hekker2011b}
{Hekker}, S., {Gilliland}, R.~L., {Elsworth}, Y., {et~al.} 2011, \mnras, 414,
  2594

\bibitem[{Hekker {et~al.}(2009)Hekker, Kallinger, Baudin, {De Ridder}, Barban,
  Carrier, Hatzes, Weiss, \& Baglin}]{Hekker09}
Hekker, S., Kallinger, T., Baudin, F., {et~al.} 2009, \aap, 506, 465

\bibitem[{{Howell} {et~al.}(2014){Howell}, {Sobeck}, {Haas}, {Still},
  {Barclay}, {Mullally}, {Troeltzsch}, {Aigrain}, {Bryson}, {Caldwell},
  {Chaplin}, {Cochran}, {Huber}, {Marcy}, {Miglio}, {Najita}, {Smith},
  {Twicken}, \& {Fortney}}]{Howell2014}
{Howell}, S.~B., {Sobeck}, C., {Haas}, M., {et~al.} 2014, \pasp, 126, 398

\bibitem[{{Huber}(2014)}]{Huber2014}
{Huber}, D. 2014, ArXiv e-prints

\bibitem[{{Huber} {et~al.}(2013){Huber}, {Carter}, {Barbieri}, {Miglio},
  {Deck}, {Fabrycky}, {Montet}, {Buchhave}, {Chaplin}, {Hekker},
  {Montalb{\'a}n}, {Sanchis-Ojeda}, {Basu}, {Bedding}, {Campante},
  {Christensen-Dalsgaard}, {Elsworth}, {Stello}, {Arentoft}, {Ford},
  {Gilliland}, {Handberg}, {Howard}, {Isaacson}, {Johnson}, {Karoff},
  {Kawaler}, {Kjeldsen}, {Latham}, {Lund}, {Lundkvist}, {Marcy}, {Metcalfe},
  {Silva Aguirre}, \& {Winn}}]{Huber2013}
{Huber}, D., {Carter}, J.~A., {Barbieri}, M., {et~al.} 2013, Science, 342, 331

\bibitem[{{Lebreton} \& {Goupil}(2014)}]{Lebreton2014}
{Lebreton}, Y. \& {Goupil}, M.-J. 2014, ArXiv e-prints

\bibitem[{{Lillo-Box} {et~al.}(2014){Lillo-Box}, {Barrado}, {Moya},
  {Montesinos}, {Montalb{\'a}n}, {Bayo}, {Barbieri}, {R{\'e}gulo}, {Mancini},
  {Bouy}, \& {Henning}}]{Lillo-Box2014}
{Lillo-Box}, J., {Barrado}, D., {Moya}, A., {et~al.} 2014, \aap, 562, A109

\bibitem[{{Maeder}(1975)}]{Maeder1975}
{Maeder}, A. 1975, \aap, 43, 61

\bibitem[{{Miglio} {et~al.}(2012){Miglio}, {Brogaard}, {Stello}, {Chaplin},
  {D'Antona}, {Montalb{\'a}n}, {Basu}, {Bressan}, {Grundahl}, {Pinsonneault},
  {Serenelli}, {Elsworth}, {Hekker}, {Kallinger}, {Mosser}, {Ventura},
  {Bonanno}, {Noels}, {Silva Aguirre}, {Szabo}, {Li}, {McCauliff}, {Middour},
  \& {Kjeldsen}}]{Miglio2012}
{Miglio}, A., {Brogaard}, K., {Stello}, D., {et~al.} 2012, \mnras, 419, 2077

\bibitem[{{Miglio} {et~al.}(2014){Miglio}, {Chaplin}, {Farmer}, {Kolb},
  {Girardi}, {Elsworth}, {Appourchaux}, \& {Handberg}}]{Miglio2014}
{Miglio}, A., {Chaplin}, W.~J., {Farmer}, R., {et~al.} 2014, \apjl, 784, L3

\bibitem[{{Miglio} {et~al.}(2013){Miglio}, {Chiappini}, {Morel}, {Barbieri},
  {Chaplin}, {Girardi}, {Montalb{\'a}n}, {Noels}, {Valentini}, {Mosser},
  {Baudin}, {Casagrande}, {Fossati}, {Silva Aguirre}, \& {Baglin}}]{Miglio2013}
{Miglio}, A., {Chiappini}, C., {Morel}, T., {et~al.} 2013, in European Physical
  Journal Web of Conferences, Vol.~43, European Physical Journal Web of
  Conferences, 3004

\bibitem[{{Montalb{\'a}n} {et~al.}(2013){Montalb{\'a}n}, {Miglio}, {Noels},
  {Dupret}, {Scuflaire}, \& {Ventura}}]{Montalban2013}
{Montalb{\'a}n}, J., {Miglio}, A., {Noels}, A., {et~al.} 2013, \apj, 766, 118

\bibitem[{{Montalb{\'a}n} {et~al.}(2012){Montalb{\'a}n}, {Miglio}, {Noels},
  {Scuflaire}, {Ventura}, \& {D'Antona}}]{Montalban2012}
{Montalb{\'a}n}, J., {Miglio}, A., {Noels}, A., {et~al.} 2012, in Red Giants as
  Probes of the Structure and Evolution of the Milky Way, ed. A.~{Miglio},
  J.~{Montalb{\'a}n}, A.~{Noels}, A.~Miglio, J.~Montalb{\'a}n, \& A.~Noels,
  ApSS Proceedings, 23

\bibitem[{{Montalb{\'a}n} \& {Noels}(2013)}]{Montalban2013a}
{Montalb{\'a}n}, J. \& {Noels}, A. 2013, in European Physical Journal Web of
  Conferences, Vol.~43, European Physical Journal Web of Conferences, 3002

\bibitem[{{Mosser} {et~al.}(2011){Mosser}, {Barban}, {Montalb{\'a}n}, {Beck},
  {Miglio}, {Belkacem}, {Goupil}, {Hekker}, {De Ridder}, {Dupret}, {Elsworth},
  {Noels}, {Baudin}, {Michel}, {Samadi}, {Auvergne}, {Baglin}, \&
  {Catala}}]{Mosser2011}
{Mosser}, B., {Barban}, C., {Montalb{\'a}n}, J., {et~al.} 2011, \aap, 532, A86

\bibitem[{{Mosser} {et~al.}(2010){Mosser}, {Belkacem}, {Goupil}, {Miglio},
  {Morel}, {Barban}, {Baudin}, {Hekker}, {Samadi}, {De Ridder}, {Weiss},
  {Auvergne}, \& {Baglin}}]{Mosser2010}
{Mosser}, B., {Belkacem}, K., {Goupil}, M.-J., {et~al.} 2010, \aap, 517, A22

\bibitem[{{Rauer} {et~al.}(2013){Rauer}, {Catala}, {Aerts}, {Appourchaux},
  {Benz}, {Brandeker}, {Christensen-Dalsgaard}, {Deleuil}, {Gizon}, {Goupil},
  {G{\"u}del}, {Janot-Pacheco}, {Mas-Hesse}, {Pagano}, {Piotto}, {Pollacco},
  {Santos}, {Smith}, {-C.}, {Su{\'a}rez}, {Szab{\'o}}, {Udry}, {Adibekyan},
  {Alibert}, {Almenara}, {Amaro-Seoane}, {Ammler-von Eiff}, {Asplund},
  {Antonello}, {Ball}, {Barnes}, {Baudin}, {Belkacem}, {Bergemann}, {Bihain},
  {Birch}, {Bonfils}, {Boisse}, {Bonomo}, {Borsa}, {Brand{\~a}o}, {Brocato},
  {Brun}, {Burleigh}, {Burston}, {Cabrera}, {Cassisi}, {Chaplin}, {Charpinet},
  {Chiappini}, {Church}, {Csizmadia}, {Cunha}, {Damasso}, {Davies}, {Deeg},
  {D{\'{\I}}az}, {Dreizler}, {Dreyer}, {Eggenberger}, {Ehrenreich},
  {Eigm{\"u}ller}, {Erikson}, {Farmer}, {Feltzing}, {de Oliveira Fialho},
  {Figueira}, {Forveille}, {Fridlund}, {Garc{\'{\i}}a}, {Giommi}, {Giuffrida},
  {Godolt}, {Gomes da Silva}, {Granzer}, {Grenfell}, {Grotsch-Noels},
  {G{\"u}nther}, {Haswell}, {Hatzes}, {H{\'e}brard}, {Hekker}, {Helled},
  {Heng}, {Jenkins}, {Johansen}, {Khodachenko}, {Kislyakova}, {Kley}, {Kolb},
  {Krivova}, {Kupka}, {Lammer}, {Lanza}, {Lebreton}, {Magrin}, {Marcos-Arenal},
  {Marrese}, {Marques}, {Martins}, {Mathis}, {Mathur}, {Messina}, {Miglio},
  {Montalban}, {Montalto}, {Monteiro}, {Moradi}, {Moravveji}, {Mordasini},
  {Morel}, {Mortier}, {Nascimbeni}, {Nelson}, {Nielsen}, {Noack}, {Norton},
  {Ofir}, {Oshagh}, {Ouazzani}, {P{\'a}pics}, {Parro}, {Petit}, {Plez},
  {Poretti}, {Quirrenbach}, {Ragazzoni}, {Raimondo}, {Rainer}, {Reese},
  {Redmer}, {Reffert}, {Rojas-Ayala}, {Roxburgh}, {Salmon}, {Santerne},
  {Schneider}, {Schou}, {Schuh}, {Schunker}, {Silva-Valio}, {Silvotti},
  {Skillen}, {Snellen}, {Sohl}, {Sousa}, {Sozzetti}, {Stello}, {Strassmeier},
  {{\v S}vanda}, {Szab{\'o}}, {Tkachenko}, {Valencia}, {van Grootel},
  {Vauclair}, {Ventura}, {Wagner}, {Walton}, {Weingrill}, {Werner}, {Wheatley},
  \& {Zwintz}}]{Rauer2013}
{Rauer}, H., {Catala}, C., {Aerts}, C., {et~al.} 2013, ArXiv e-prints

\bibitem[{{Reimers}(1975)}]{Reimers1975a}
{Reimers}, D. 1975, M{\'e}moires of the Soci{\'e}t{\'e} Royale des Sciences de
  Li{\`e}ge, 8, 369

\bibitem[{{Ricker}(2014)}]{Ricker2014}
{Ricker}, G.~R. 2014, Journal of the American Association of Variable Star
  Observers (JAAVSO), 42, 234

\bibitem[{{Rodrigues} {et~al.}(2014){Rodrigues}, {Girardi}, {Miglio},
  {Bossini}, {Chaplin}, {Girardi}, {Montalb{\'a}n}, {Valentini}, {Mosser},
  {Baudin}, {Casagrande}, {Fossati}, {Aguirre}, \& {Baglin}}]{Rodrigues2014}
{Rodrigues}, T., {Girardi}, L., {Miglio}, A., {et~al.} 2014, \mnras, accepted

\bibitem[{Roxburgh \& Vorontsov(2003)}]{Roxburgh03}
Roxburgh, I.~W. \& Vorontsov, S.~V. 2003, \aap, 411, 215

\bibitem[{{Stello} {et~al.}(2013){Stello}, {Huber}, {Bedding}, {Benomar},
  {Bildsten}, {Elsworth}, {Gilliland}, {Mosser}, {Paxton}, \&
  {White}}]{Stello2013}
{Stello}, D., {Huber}, D., {Bedding}, T.~R., {et~al.} 2013, \apjl, 765, L41

\bibitem[{{White} {et~al.}(2011){White}, {Bedding}, {Stello},
  {Christensen-Dalsgaard}, {Huber}, \& {Kjeldsen}}]{White2011}
{White}, T.~R., {Bedding}, T.~R., {Stello}, D., {et~al.} 2011, \apj, 743, 161

\end{thebibliography}
\end{document}